\documentclass[aps,prd,twocolumn,groupedaddress,showpacs,fleqn]{revtex4}
\usepackage{graphicx}
\usepackage{epsfig}

\usepackage{mathrsfs}
\usepackage{amssymb}
\usepackage[intlimits]{amsmath}

\begin{document}
\numberwithin{equation}{section}

\title{Energy conservation drives the expansion of the universe}


\author{J. M. Greben}

\email[]{jmgreben@gmail.com}
\altaffiliation {Some of this work was carried out while the author was principal scientist at the CSIR (Pretoria, South Africa) before his retirement in 2012}


\begin{abstract}
We develop a cosmological theory in which the evolution of the universe is controlled by the cosmological constant and dominated by the associated vacuum energy. The universe starts as a classical de Sitter space with an infinite effective vacuum energy density, 
which decreases subsequently like $t^{-3}$. The corresponding Friedmann-Robertson-Walker (FRW) scale factor also decreases over time, showing that the common assumption that it describes the expansion of  the universe is incorrect and should be abandoned. Instead, the (cubic) expansion of the universe is needed to satisfy energy conservation. Once the vacuum energy density has decreased to the Planck level the first elementary particles can be created through a direct conversion of vacuum energy. After this epoch, the enormous kinetic energy enables a quick magnification of the number of particles through ordinary production processes in tandem with the expansion of space.


The dominance of  vacuum (dark) energy is supposed to persist into later epochs, which enables a perturbative treatment of matter and radiation leading to linear equations, which replace the usual FRW equations. The presence of matter changes the vacuum metric, inducing a secondary matter term which might explain the phenomenon of dark matter. Together with a similar induced radiation term, it provides a possible explanation for the recent acceleration of the expansion of the universe. The theory unifies particle physics and cosmology by expressing particle physics units in terms of the gravitational and cosmological constant. This relationship also explains a number of numerical coincidences which have long puzzled physicists.

\end{abstract}

\pacs{98.80.-k, 98.80.Bp, 98.80.Cq}
\maketitle
\def\thesection{\arabic{section}}
\section{Introduction}
\label{sec:Introduction}
Conservation of energy is (one of) the most basic law(s) in Nature. However, in general relativity the satisfaction of this law is not so straightforward. The problem lies in the time dependence of the metric and the expansion of space. General relativity is expressed in terms of energy density, while energy conservation is expressed in terms of energy. To relate the two entities,  one needs to integrate the density over space. This raises two points: how do we account for the changing metric in this spatial integral; and how do we account for the expansion of space and the consequential increase in the volume of the spatial integral? These questions are further complicated by the fact that the expansion itself has so far been seen as a consequence of the metric, highlighting the complexity of this non-linear web of relationships. The presence of different forms of energy in the universe further complicates the situation.

To minimize these complexities in the analysis of energy conservation, it makes sense to consider the simplest non-trivial space, namely a vacuum space with a positive cosmological constant. This de Sitter space is completely homogeneous and only contains one form of energy: vacuum energy. Furthermore, the energy tensor is itself expressed in terms of the metric tensor, so this is a fully self-contained problem which admits an exact non-linear solution of the Einstein equations.

The metric of this space has a particularly elegant form if it is expressed in conformal time, as it is proportional to $1/t^2$. The singular point $t=0$ can then be identified with the big bang. If we now compute the energy in an arbitrary volume $V$, then the energy density within the energy integral is multiplied with a metric factor that behaves like $t^{-3}$. So, the product, which can be called the effective energy density, is infinite at the big bang and decreases rapidly afterwards.

This simple analysis already yields a number of further insights. First, since the physics should not depend fundamentally on the metric used,
it is of interest to convert the conformal solution to the Friedmann-Robertson-Walker (FRW) metric which is more common in cosmology. This metric is expressed in terms of a scale factor which is proportional to the proper distance and which is supposed to describe the expansion of the universe. However, after converting the conformal solution one finds that the FRW scale factor decreases after the big bang, so it does not describe the expansion of the universe. Hence, the main assumption in the FRW framework, which underlies much of the standard $\Lambda$CDM model of cosmology, is incorrect. We must conclude that general relativity does not control or predict the expansion of the universe, but that the expansion is a consequence of the demand of energy conservation.

A second insight is that conformal time, rather than the proper time used in FRW, should be used in cosmology. In proper time the big bang lies in the infinite past. This may be the actual experience of a (proper) observer, but does not match the time frame common in cosmology, where the big bang happens at a definite time (usually $t=0$) and the age of the universe equals about 14 billion years, as is the case in conformal time.

A third insight is that this de Sitter space has a number of properties which makes it ideal as a model for the big bang and the early universe. These include the fact that the classical description allows for the instantaneous transition from nothingness to a fully specified state. Also, the entropy of this classical state is zero as the state is unique ($N=1$ in $S=k \ln \,N)$. Penrose \cite{PenroseCQC} has repeatedly stressed that the extreme smallness (or vanishing) of the entropy and the vanishing of the Weyl tensor (or conformal tensor) are essential at the singularity of the big bang. It is easy to prove that the Weyl tensor is indeed zero in this de Sitter universe. Zero entropy is also the natural starting point for the thermodynamic evolution of the universe. That the initial density is infinite is also what one would expect if the big bang happens at a definite point in time. Finally, the vacuum space with a positive cosmological constant is possibly the simplest non-trivial state of the universe. With the complexity of the universe increasing all the time, this may well be the optimal choice for the initial state.

However, to accept such a classical description of the early universe, a number of conditions must be met. First, the cosmological constant must be a classical constant; second, any quantum fluctuations in the early universe should have no effect on the state (vector); and third, the initial classical state should  persist for a finite time so that the (classical) solution gives a complete description of this early epoch. We will show below that these conditions are indeed satisfied.

But first we return to the original issue: energy conservation and general relativity. We already mentioned that the effective energy density in the vacuum universe behaves like $t^{-3}$. So to conserve energy in a volume $V$, this volume must expand like $t^{3}$, corresponding to a linear expansion in all spatial directions. However, as the volume cannot extend into the existing surrounding space, this expansion must take place throughout the whole universe.  Since the de Sitter space is completely homogeneous, this presents no problem and the expansion simply implies a change of scale in the whole universe, which can be represented by a conformal scale factor. This expansion is a consequence of energy conservation and does not interfere with the derivations in general relativity. However, when calculating energy integrals (which must be compared at different times) the change in volume ensures conservation of energy.


Although the expansion guarantees the conservation of (vacuum) energy in the de Sitter space, this argument fails at the big bang itself. Whether energy conservation can be demanded at the very moment of creation is more of a philosophical question, as the big bang lies in the infinite past for a proper observer and also marks the beginning of time.
In this connection, it is worthwhile to recall an argument by Stenger \cite{Stenger} who suggests a possible explanation for the creation of energy in the big bang which is in accordance with energy conservation.
This argument is based on a generalization of the proof of energy conservation in QFT. By carrying out a spatial integration over the (vanishing) divergence of the energy-momentum tensor in QFT, one obtains an identity that expresses the conservation of energy. In general relativity this divergence is replaced by its covariant counterpart, which confirms that the proper generalisation of the energy integral involves the effective energy density. But there is also an additional (negative) pressure term. This gives the overall identity the form of the first law of thermodynamics and led Stenger \cite{Stenger} to argue that the creation of energy in the big bang is due to an adiabatic expansion. The creation of energy at the big bang is then a consequence of the work done on the system by the negative pressure (the ultimate free lunch). This could thus serve as an explanation for the (instantaneous) creation of energy in the big bang.

After the big bang we do not need a pressure term to create the vacuum energy and one can show that the energy and pressure term are conserved separately if the volume increases as indicated. The first law of thermodynamics is then trivially satisfied. In this way the role of the odd pressure term has been limited to the very moment of creation. This is of some importance, as the interpretation of this negative pressure term is unclear. As Taylor \cite{Taylor} states: there is no known fundamental substance with this property. Chan \cite{Chan} is even willing to abandon energy conservation in order to eliminate the pressure term. By decoupling the energy and pressure term after the big bang, we have found another way to re-focus on the energy term, thereby also facilitating the incorporation of QFT in cosmology.

The early classical epoch is followed by an epoch, where the (first) elementary particles are created. To describe this unique process, where vacuum energy is converted into particles, we rely on an extension of QFT in which the elementary particles of the Standard Model are modelled as finite Planckian systems \cite{QuarkDressing}, \cite{QuarkSpatial}. In the calculation of the energy of these particles, the inclusion of general relativity is essential, as well as a positive energy term to compensate for the negative potential energy in the QFT state. This energy term must have an enormous Planckian density, which can realized by a very young finite spherical vacuum universe. These properties, which it shares with the early universe, make this model of elementary particles ideally suited to handle the unique particle creation process in the early universe.

The initial creation of elementary particles can only take place after the effective vacuum energy density has dropped to the Planck level ($\sim G^{-2}$), at which point it matches the internal energy density of elementary particles. This explains why the classical epoch persists for a finite time (for about $10^{-23}$s).
The potential energy is then converted into kinetic energy (the mass is negligible), making these particles extremely energetic with momenta in the Planck energy range.

After this unique creation epoch, particles can only be created or annihilated by standard collision processes or decays. The enormous kinetic energy of the initial particles will quickly be distributed over subsequent generations of elementary particles, whose creation becomes possible in tandem with the increase of the available space. When the kinetic energy becomes comparable to typical particle masses, the annihilation of particles and anti-particles takes over from the earlier creation processes. This should eventually result in a universe that is dominated by electron and nucleons, rather than by positrons and anti-nucleons. An explanation for this asymmetry could lie in the observed asymmetry in the gauge interactions (so far these effects seem to be too small) or lie in the competition between different processes in these early stages of the universe and be of a more statistical origin. To analyze the latter possibilities the BS-QFT theory has to be further developed for leptons and heavier quarks, as in its current form it only gives accurate answers for light quarks \cite{QuarkSpatial} .

The description of later - but still early - phases of the evolution of the universe, such as the nucleosynthesis and recombination era, are expected to be more sensitive to the specific physical conditions present in different eras, rather than to the description of the first seconds of the universe and will not be discussed here.

In the later phase of the universe, when the composition of the universe in terms of matter and radiation has more or less stabilised, the continued dominance of vacuum (dark) energy, as seen in recent observations \cite{Kowalski}, allows us to treat the matter and radiation components in the Einstein equations perturbatively. The resulting linear differential equations then replace the usual FRW equations. This treatment also reveals that the linear expansion of the early universe resumes in this phase, and only requires adjustments when the composition of the universe changes globally. This linearity contributes highly to the elegance and functionality of the cosmological theory. It should not be confused with the linear scale factor occurring in the Milne universe \cite{Milne}, as this is an FRW scale factor defined in terms of proper time.

The perturbative approach also makes it possible to refine the cosmological principle by taking into account the non-homogeneity pointlike nature of matter in the universe. Thus we are able to include contributions of spatial derivatives in the Einstein equations, making its application more realistic. This also enables us to make a connection between global cosmological theory and local gravitational phenomena and analyze how - and why - local systems remain stable under the expansion of the universe. The similarities between this perturbative theory and weak gravity theory can also be exploited.

The current approach also allows an analysis of the global energy content of the universe. The modified vacuum metric gives rise to an extra matter term, which equals 9/4 times the ordinary matter term, a factor which is likely to increase after further refinement of the theory. This term could be an explanation for the dark matter phenomenon, although the connection between this global result and the local impact of dark matter in galaxies must still be demonstrated. There is also a cross term between the vacuum and radiation energy which reduces the basic radiation term. Unfortunately, there is accidental ambiguity in its determination which needs to be resolved. The acceleration of the expansion of the universe, which in this theory can no longer be attributed to the presence of the cosmological constant, can now be attributed to these cross terms, as the global conversion of matter into radiation reduces the induced energy terms, necessitating an acceleration of the expansion.

The outline of the paper is as follows. In Sec.\ \ref{sec:Analysis} we discuss the solutions of the Einstein equations in the de Sitter space under the conformal and the FRW metric.
This is followed by a short section on the cosmological constant, which plays such an important role in our cosmological theory. In Sec.\ \ref{sec:Creation} we describe the first creation of elementary particles. In Sec.\ \ref{sec:Matter_Rad} we develop the perturbative approach to matter and radiation components, deriving the linear equations of motion which replace the FRW equations. In Sec.\ \ref{sec:Balance} follows a discussion of the different contributions to the total energy. In Sec.\ \ref{sec:Big_Numbers} we present an explanation of a number of numerical coincidences which have puzzled physicists like Eddington, Dirac and Weinberg over the last century. Finally, in Sec.\ \ref{sec:Summary} we summarize the main achievements of the new cosmological approach and discuss the way forward.



\section{Energy conservation in the vacuum universe}
\label{sec:Analysis}
The analysis of energy conservation is carried out in the flat de Sitter space \cite{Sitter} with a positive cosmological constant $\Lambda$. This space is completely homogeneous, so that a discussion of a finite volume is representative of the whole universe, facilitating the discussion of energy conservation. The energy-momentum tensor for this space equals:
\begin{equation}
 \label{eq:Tuv_vacuum}
 T_{\mu\nu}(x)=-\epsilon g_{\mu\nu}(x),
\end{equation}
where $\epsilon$ is the vacuum energy density. This density is related to the cosmological constant $\Lambda$ by:
\begin{equation}
 \label{eq:Vac_Energy}
 \epsilon=\Lambda/8\pi G,
\end{equation}
where $G$ is the gravitational constant. In cosmology one usually assumes that space is homogenous and isotropic (the cosmological principle), so that the energy-momentum tensor can be parameterized as follows \cite {Copeland}:
\begin{equation}
 \label{eq:T_general}
  T_{\mu}^{\nu}(x)=\delta_{\mu\nu}\left[-\rho(t)\delta_{\mu 0}+p(t)\bar{\delta}_{\mu 0}\right],
\end{equation}
 where $\rho(t)$ is the average energy density and $p(t)$ is the average pressure density. This parametrization is exact in the vacuum case  with both $\rho(t)=\epsilon$ and $p(t)=-\epsilon$ constant. Note that the vacuum pressure is negative.

We now analyze the solutions of the Einstein equations in this vacuum space for two different metrics. The FRW metric, which is expressed in terms of the scale factor $a(\tau)$, is given by:
\begin{eqnarray}
 \label{eq:RW_metric}
 \begin{aligned}
 ds^2=g_{\mu\nu}dx^{\mu} dx^{\nu} = -d\tau^2 + a^2(\tau)\,d\sigma^2,
 \end{aligned}
\end{eqnarray}
while the conformal metric is defined in terms of a conformal factor $g(t)$ that multiplies the Minkowski metric:
\begin{eqnarray}
 \label{eq:Conformal_metric}
 ds^2=g(t)\eta_{\mu\nu}dx^{\mu} dx^{\nu}= g(t)\left(-dt^2+d \sigma^2\right).
\end{eqnarray}
The spatial line element $d\sigma$ is given by:
\begin{equation}
 \label{eq:dsigma2}
d\sigma^2=dr^2+r^2\left(d\theta^2+\sin^2\theta\, d\phi^2\right),
\end{equation}
where we assumed that space is flat, as is generally accepted by now (the curvature parameter $\Lambda_k=-0.0027\pm 0.0038$ according to the summary given in \cite{Bennett}). To distinguish the proper time used in the FRW metric from the conformal time $t$ used in the conformal metric, we indicate the former by the symbol $\tau$.

In this de Sitter space the solutions of the Einstein equations for the FRW scale factor are:
\begin{equation}
 \label{eq:scale_factor}
 a(\tau)\sim \exp(\pm H \tau );H=\left(8\pi G \epsilon /3\right)^{1/2}=(\Lambda/3)^{1/2}.
\end{equation}
If $a(\tau)$ describes the expansion of the universe, as is usually assumed, then the positive sign must be chosen. Under the conformal metric we obtain:
\begin{eqnarray}
 \label{eq:Classical_Solution}
g(t)=t_s^2/t^2;t_s=(3/8G\pi\epsilon)^{1/2}=H^{-1}.
\end{eqnarray}
This elegant time-symmetric solution has received little attention in the literature, although it was mentioned by Ibison \cite{Ibison} in a review article on metrics. The metric is singular at $t=0$, which point is identified with the big bang. The conformal solution is essentially unique, as the two branches ($t>0$ and $t<0$) are physically equivalent, and choosing $t>0$ is just a matter of convention.


The conformal metric decreases after the big bang and since physics should not depend in a fundamental way on the choice of metric, the FRW scale factor should decrease as well. So the exponentially decreasing solution in Eq.\ (\ref{eq:scale_factor}) is the physical correct one, implying that the standard hypothesis that the scale factor describes the expansion of the universe is incorrect. Or phrased differently, the proper distance does not reflect the expansion of the universe, contrary to what is usually assumed \cite{MTW}.

One could object that inversely, the common exponentially increasing FRW solution should lead to a corresponding increasing conformal solution. However, if one converts this FRW solution to the conformal metric then one finds that the big bang corresponds to $\tau=\infty$, so that time must run backwards and the FRW scale factor again decreases after the big bang. This failure of the FRW formulation to identify the right physical solution on its own might be attributed to the fact that the exponential function does not have a singular point which can be identified with the big bang. However, if we accept that the scale factor modifies the (vacuum) energy density and thus should be maximal near the big bang - as we argue in Eq.\ (\ref{eq:effective_density}) - then one would immediately be able to conclude that the decreasing solution is the correct one.


We now establish the explicit relationship between the two metrics:
\begin{equation}
\label{eq:model_conversion}
\frac{t_s}{t}dt=d\tau \rightarrow  \tau =t_s+t_s \ln\left(\frac{t}{t_s}\right),
\end{equation}
where we choose the integration constant such that $t$ and $\tau$ share the same reference point $t_s$. This equation was already mentioned by Milne \cite{MilneFD}, though in a different context. He identifies $t_s$ (which he calls $t_0$) with the age of the universe, and indeed we will find in Sec.\ \ref{sec:CC} that $t_s\approx 13.9\times 10^9$ years, in good agreement with the accepted age of the universe. The reason that the age of the universe appears constant will also be explained in Sec.\ \ref{sec:CC}.

If we use this equation to convert the conformal solution to FRW form we get - as expected - the decreasing exponential:
\begin{eqnarray}
 \label{eq:RW_conf}
a(\tau) =g^{1/2}(t)=\exp\left(-\frac{\tau-t_s}{t_s}\right).
\end{eqnarray}
Both $g(t)$ and $a(\tau)$ equal unity when $t$ and $\tau$ correspond to the age of the universe $t_s$, further justifying the choice of the normalization in Eq.\ (\ref{eq:model_conversion}).

The big bang at $t=0$ now corresponds to a proper time of $\tau\rightarrow-\infty$, implying that the big bang is unreachable for a proper observer, but also establishing that conformal time is the natural time variable to use in cosmology as it starts at $t=0$. The conformal metric has numerous other advantages. As Penrose notes: conformal rescaling (in this case of the Minkowski metric) leaves the causal relationships between points unaffected \cite{PenroseCQC}. Also, the geodesic has the same form as in Minkowski space, except for the limitation that $t\geq0$ or $t\leq0$. Another advantage is that matter and radiation can be incorporated in a perturbative manner in the conformal metric, revealing similarities with weak gravity, and thus access to its well-developed techniques (see Sec.\ \ref{sec:Matter_Rad}).

We now analyze the conservation of energy in our de Sitter space. The energy contained in an arbitrary volume $V$ is given by:
\begin{eqnarray}
 \label{eq:Energy_V}
 \begin{aligned}
 E_V=-\int\limits_{V}d^3x |g_{ij}|^{1/2}\,T^0_0=\int\limits_{V}d^3x |g_{ij}|^{1/2}\,\epsilon,
 \end{aligned}
\end{eqnarray}
where $|g_{ij}|$ is the spatial part of the metric determinant, which equals $a^3(\tau)$ in the FRW case and $g^{3/2}(t)$ in the conformal case.
It is now natural to call its product with $\epsilon$ the effective (vacuum) energy density:
\begin{equation}
 \label{eq:effective_density}
 \epsilon^{eff}(t)=g^{3/2}(t)\epsilon=\frac{t^3_s}{t^3}\epsilon.
\end{equation}
This effective energy density becomes large (infinite) near (at) the big bang. This is what one would expect, and gives the physical explanation why $a(\tau)$ or $g(t)$ should be maximal near the big bang.

The energy $E_V$ contained in the volume $V$ now decreases like $1/t^3$ in conformal time and like $\exp (-3H\tau)$ in proper time. Hence, in both cases the energy is not conserved; as would have been the case in the original FRW formulation where $E_V$ would increase exponentially.

But there is a way to restore energy conservation. By allowing the volume $V$ to increase with time:
  \begin{equation}
 \label{eq:V_t}
 V \rightarrow V(t)=V_s\frac{t^3}{t^3_s},
\end{equation}
one can compensate for the decrease in energy density, thereby ensuring that $E_V$ is conserved. Here $V_s$ is the volume at the reference time $t_s$. Hence, the expansion of the universe is a consequence of energy conservation, instead of being determined by the metric.

Does this mean that energy is not conserved automatically in general relativity? To answer this question we review how energy conservation is proved in QFT and extend this proof to general relativity. This will also provide a possible explanation for the very origin of energy in the big bang.

In QFT energy conservation is proved using the identity:
\begin{eqnarray}
 \label{eq:Divergence}
 \partial_{\mu} T^{\mu}_{\:\nu} =0\;(\textrm{sum over }\mu \textrm{ understood}),
\end{eqnarray}
and integrating this over space. In general relativity this divergence is replaced by the covariant divergence, leading to two extra terms:
\begin{eqnarray}
 \label{eq:CovDer}
 \nabla_{\mu} T^{\mu}_{\:\nu} =\partial_{\mu} T^{\mu}_{\:\nu} -\Gamma^{\lambda}_{\mu\nu}T^{\mu}_{\:\lambda}+\Gamma^{\mu}_{\mu\lambda}T^{\lambda}_{\:\nu}=0.
\end{eqnarray}
Note that this identity is trivially satisfied for the vacuum energy density even if the metric is modified by matter and radiation, as $\nabla^{\mu} g_{\mu\nu} =0$ is one of the basic identities in general relativity. In our de Sitter space the connection coefficients are given by:
\begin{eqnarray}
 \label{eq:con_coeff1}
  \begin{aligned}
&\Gamma^{\lambda}_{\mu\nu}=-\frac{1}{t}\left(\delta_{\mu 0}\delta_{\nu \lambda}+\delta_{\nu 0}\delta_{\mu \lambda}-\eta^{\lambda 0}\eta_{\mu\nu}\right),
\\
&\sum_{\mu}\Gamma^{\mu}_{\mu\lambda}=-\frac{4}{t}\delta_{\lambda 0}.
 \end{aligned}
\end{eqnarray}
We now carry out the spatial integral over the covariant divergence for $\nu=0$:
\begin{eqnarray}
 \label{eq:Energy_relation}
 \begin{aligned}
 0=&-\int\limits_{V}d^3x \frac{t^3_s}{t^3}\;\partial_0 T^0_{\:0}+\int\limits_{V}d^3x \frac{t^3_s}{t^3}\;\nabla_iT^i_{\:0}+
 \\
 -&\frac{1}{t}\int\limits_{V}d^3x \frac{t^3_s}{t^3}T^{\mu}_{\:\mu}+\frac{4}{t}\int\limits_{V}d^3x \frac{t^3_s}{t^3} T^0_{\:0},
 \end{aligned}
\end{eqnarray}
where we used the conventions $\partial_i=-(\vec{\nabla})_i$.
We can rewrite this as follows:
\begin{eqnarray}
 \label{eq:EnergThermo}
 \begin{aligned}
- &\partial_0\int\limits_{V}d^3x \frac{t^3_s}{t^3}T^0_{\:0}+\left(\frac{1}{3} T^{i}_{\:i}\right)\,\partial_0\int\limits_{V}d^3x
 \frac{t^3_s}{t^3}=
 \\
 &= \partial_0E_V+p\partial_0 \hat{V}=0,
 \end{aligned}
\end{eqnarray}
where we used $T^i_{\:i}=3p$  and introduced the proper volume $\hat{V}$, which equals the volume $V$ times the spatial metric factor:
\begin{equation}
\label{eq:Proper_volume}
 \hat{V}=g^{3/2}(t)V=\frac{t^3_s}{t^3}V.
\end{equation}
In deriving Eq.\ (\ref{eq:EnergThermo}) we have dropped the momentum term $T^i_{\:0}$ as diagonal elements are absent in the cosmological parametrization of $T_{\mu}^{\;\nu}$ in Eq.\ (\ref{eq:T_general}). Even in the general case (like QFT) it is considered negligible, as:
 \begin{eqnarray}
 \label{eq:Surface_momentum_term}
 \begin{aligned}
 \int\limits_{V}d^3x \frac{t^3_s}{t^3} \nabla_i T^i_{\:0}=\int\limits_{S}dS \frac{t^3_s}{t^3} n_i T^i_{\:0},
 \end{aligned}
\end{eqnarray}
represents the difference between the momentum influx and outflow, which should be negligible for a large enough volume $V$.

In QFT the pressure term in Eq.\ (\ref{eq:EnergThermo}) is absent, so that $E_V$ is conserved by itself, in agreement with the condition imposed earlier.

With the additional pressure term, Eq.\ (\ref{eq:EnergThermo}) takes on the character of the first law of thermodynamics (the entropy term is absent as the entropy is zero). On this basis Stenger \cite{Stenger} argues that the creation of the initial energy in the universe can be generated by an adiabatic expansion of the universe thanks to the negative pressure term (the ultimate free lunch). This would be a rather exceptional phenomenon, as normally an adiabatic expansion entails a reduction of the internal energy caused by positive pressure. In thermodynamics a negative pressure term is considered to act from the outside, which also seems odd, as there is no outside universe. As we recalled in the introduction, other authors also had problems with this odd pressure term (see \cite {Taylor} and \cite {Chan}).

The interpretation of the pressure term becomes more acceptable once we note that the proper volume $\hat{V}$ does not increase, but decreases like $1/t^3$, where the time dependence of $V$ is ignored as it is due to scaling and should be excluded from dynamical derivations. We then find that the pressure term $p\partial_0 \hat{V}$ is positive, so that the internal energy decreases in time and the pressure works from the inside. So energy is conserved in the form of Eq.\ (\ref{eq:EnergThermo}); but to explain the expansion of the universe one needs to demand that $E_V$ is conserved.

This condition also ensures the conservation of the pressure term. Hence, the thermodynamic equation is automatically satisfied. By having separate conservation laws, we have essentially decoupled the pressure term from the energy term, so that we can focus on the latter. This also facilitates the incorporation of quantum physics in cosmology as the pressure term is absent in QFT.


Other authors have also focused on energy integrals, but follow different approaches to enforce energy conservation. For example, Landau and Lifshitz \cite{LL} and MTW \cite{MTW} introduce effective energy-momentum operators which are constructed such that their integral is conserved. This requires certain approximations which are not needed in our approach. Arnowitt, Deser and Misner \cite{ADM} transform the energy integral into a surface integral, which takes us further away from our goal to embed QFT in the cosmological theory. Also, they assume that the metric approaches the Minkowki metric asymptotically, which is not true in our theory.
We must conclude that - despite all these efforts - it has proved impossible to define an exact energy operator that ensures energy conservation without allowing space to expand.

The separate conservation of $E_V$ and $\hat{V}$ can clearly not explain the original creation of energy in the big bang, as energy and space did not exist beforehand. So, this might be the one occasion where Stenger's proposal \cite{Stenger} does apply. First, we note that the continuity equation, Eq.\ (\ref{eq:EnergThermo}), can be replaced by the discrete equation:
\begin{equation}
\label{eq:BB_EnergThermo}
   E_V+p \hat{V}=0,
\end{equation}
which is valid as $E_V=\epsilon \hat{V}$ and $p=-\epsilon$. This means that the transition from $E_V=0$ and $\hat{V}=0$ to the de Sitter universe does not have to be continuous, but can have an instantaneous character. The fact that in this singular case the pressure term might be associated with some outside space also seems less objectionable. Speculation about a possible cause for this transition falls outside the scope of this paper.
After this diversion, let us go back to the expansion of the universe, which can be characterised by a linear (conformal) scale factor:
\begin{equation}
\label{eq:a_conformal}
  a_{conf}(t)=\frac{t}{t_s}.
\end{equation}
We will see that this linear behavior continues in the presence of matter and radiation except for (small) adjustments when the composition of the universe changes.

One can now define co-moving spatial variables:
 \begin{equation}
\label{eq:co_moving}
\tilde{\vec{x}}=\frac{t_s}{t}\vec{x} \leftrightarrow \vec{x}=\frac{t}{t_s}\tilde{\vec{x}},
\end{equation}
together with the co-moving time variable $\tilde{t}$:
\begin{equation}
\label{eq:co_moving_time}
  \tilde{t}=\frac{t_s}{t}t=t_s,
\end{equation}
which can be identified with the age of the universe. The constancy of $\tilde{t}$ does not mean that time stands still for a co-moving observer, as:
\begin{equation}
\label{eq:dtau}
  d\tilde{t}=\frac{t_s}{t}\; dt=d\tau.
\end{equation}
Hence, the variation in $\tilde{t}$ is actually identical to the variation in proper time. This simple identity also illustrates the difference between a dynamical variable like $t$ or $\tau$, and a non-dynamical entity, like the scale factor $t_s/t$.



\section{The cosmological constant}
\label{sec:CC}
A basic assumption in the previous section has been that the cosmological constant is a classical constant. To justify this assumption we have to go a bit deeper into the history of the cosmological constant.

The cosmological constant has a checkered history, with Einstein calling it his worst blunder and MTW calling it a great mistake in their basic text on cosmology (\cite{MTW}, p. 411). Since then it has been taken more serious because of its possible link to dark energy, but this also has led to controversy.

The controversy, known as  the cosmological constant problem \cite{WeinbergCC}, arises because many theorists seek its origin in quantum physics. First Zel'dovich \cite{Zeldovich} assumed that it was due to vacuum polarization; then Weinberg \cite{WeinbergCC} and Caroll  \cite{CarollRR} considered it to be a consequence of the quantum vacuum Casimir effect, although Jaffe later showed that in QFT this effect does not require vacuum energy \cite{Jaffe}. Since these quantum calculations have led to ridiculously large values of the cosmological constant (between 40 and 120 orders of magnitude too large), it is clear that they are flawed.


The problem lies in the fact that QFT is developed from a particle perspective, without giving equal weight to the anti-particle perspective. In scattering processes the effect of this bias is hidden or can be countered by ad hoc prescriptions, and thus has largely gone unnoticed. But in the case of the expectation values of one-variable operators and in the case  of the equations of motion of the quantum fields, it leads to unphysical answers \cite{CC problem}. The symmetry between particles and anti-particles must be restored by extending the algebra of creation and annihilation operators with the so-called $\mathbb{R}$-product \cite{Ordering}. After this extension, the quantum contributions to the cosmological constant are non-existent and the classical nature of the cosmological constant is confirmed.

The cosmological constant plays a central role in our theory. The injection of energy at the big bang is attributed to its presence, as is the creation of particles from vacuum energy in the creation epoch, as we see in the next section. It, or rather the vacuum energy which it implies, also dominates the energy content of the universe in subsequent periods. Hence, the cosmological constant is the common denominator over all periods of the universe, though having different roles at different times. We will even show in Sec.\ \ref{sec:Creation} that - together with the gravitational constant - it defines the units in particle physics. To quote Kepler in his \emph{Harmonices Mundi}: Nature is simple and uses one cause for many effects \cite{Kepler}.

Since this fundamental constant is of such importance, it is also important to know its value. As a constant of nature, its value cannot be derived from any deeper theory and must be determined from experiment or observation. Its approximate value was previously determined in \cite{FOS_Greben} and \cite{GRcosmology} by using the conformal scale factor in Eq.\ (\ref{eq:a_conformal}) to analyze supernovae data. We obtained an approximate value of $3.97\times10^{-47}\textrm{GeV}^4$ for the vacuum energy density $\epsilon$, which translates into a value of $6.69\times10^{-84} \textrm{GeV}^2$ for the cosmological constant $\Lambda$.

These constants lead to the following estimates of $t_s\approx 13.9\times 10^9$ years, which agrees well with the empirical value of the age of the universe: $13.77\pm 0.06\times 10^9$ years \cite{Bennett}; and the Hubble constant $H_0=1/t_s$ (see Eq.\ (\ref{eq:t_s})), which works out to 69.8 km/sec/Mpc in the usual Hubble constant units. This latter estimate agrees very well with the recent values given by Freedman et al.\ \cite{Freedman2019} 69.8$\pm$1.9 km/sec/Mpc and Bennett et al. \cite{Bennett} 70.0$\pm$2.2 km/sec/Mpc; which values are intermediate between those obtained in the SHOES project of Riess et al. \cite{Riess2011}: 73.8$\pm$2.4 km/sec/Mpc and the Planck based $\Lambda$CDM value of $67.4\pm 0.5$ km /sec/Mpc \cite{PCP13}.

\section{Creation of matter in the early universe}
\label{sec:Creation}
In Sec.\ \ref{sec:Analysis} we have seen that the vacuum universe with a positive cosmological constant and controlled by classical general relativity, can explain many properties of the early universe and the big bang. During this initial classical epoch quantum fluctuations have no effect as they cannot lead to real physical particles and cannot change the state of the universe. Also, virtual quantum contributions to the vacuum energy are absent as argued in the earlier discussion of the cosmological constant. So, calling this phase classical does not mean that quantum laws do not apply, but rather that they do not influence the state of the universe, with the latter fully described by classical general relativity.

Since the current universe is filled with particles and radiation, the question is: how did this initial empty universe acquire its content and transform into the current complex universe? In order to describe this transition we make use of a bound-state (BS) implementation of QFT (\cite{QuarkDressing}, \cite{QuarkSpatial}), where quarks and leptons are described as finite particles with an internal structure.

The standard formulation of QFT describes the time evolution of particles and their scattering using methods like the Feynman path integral. Instead, the BS-formulation of QFT describes the stationary states of single particles by solving the coupled field equations. These equations imply that a bare (i.e. massless and pointlike) elementary particle acts as a source of local $\textrm{SU}_{\textrm{n}}$ and $\textrm{U}_{\textrm{1}}$-gauge fields. Quantization is accomplished by expanding these fields in terms of the creation and annihilation operators of the fermionic source field. An exact (infinite) operator solution of these coupled field equations can be constructed, which in turn can be used to reduce them to a finite set of coupled non-linear differential equations for the internal radial spinor wave functions and gauge fields. Its solution describes a spherical system that is absolutely confined, providing an intuitively pleasing model for a (finite) elementary particle without the need for introducing a deeper layer of elementary particles (preons). The inclusion of general relativity is then needed to stabilize the bound system and to set the internal energy and radial scale (Planck energy and Planck length, respectively). The incorporation of general relativity in the QFT framework is facilitated by the c-number character of the reduced field equations.

To ensure that these particle solutions have a non-negative energy/mass, a positive energy term is needed to counter the internal negative QFT potential energy. The energy density of this base term must then be of $\textrm{O}(G^{-2}) \approx 2.22 \times 10^{76}\, \textrm{GeV}^4$, which is so large that only a very young - and thus still classical (and reversible) - vacuum universe can supply it. So instead of a bound state built on the masses of its constituents, as is the case for nuclei and atoms, the energy basis for finite elementary particles is a spherical vacuum universe whose expansion stops as soon as it matches the internal energy density and size of the QFT state. As we discuss at the end of this section, the overall mass of these systems turns out to be of the order of MeV, even though this requires an almost perfect cancellation of the positive and negative $G^{-1/2}$ contributions to the overall energy.

Because of the role of this vacuum universe in the creation of elementary particles, the BS-QFT theory is an ideal tool for describing the creation of (the first) particles in the universe. Since particles are absent in the original vacuum universe, we need a novel mechanism for creating particles without having to rely on the usual scattering or decay processes, as the latter require the pre-existence of particles. The spontaneous creation of a particular elementary particle in the vacuum universe is assumed to become possible as soon as the vacuum energy density in the universe has decreased to the internal energy density that characterizes the elementary particle. This creation process is then driven by the negative potential energy associated with the QFT state.

The number of spherical elementary particles that can be created in a given co-moving volume is limited by the available space and the constraint of energy conservation. Relevant packing fractions have been extensively studied: e.g. dense sphere packing yields a factor $0.74$; loose random packing yields a factor 0.56; while ellipsoidal packing yields a factor 0.77 \cite{packing}. Since particles are created together with their anti-particle, random ellipsoidal packing might be more appropriate, giving an overall factor of about 0.45.

However, whether such a large fraction of space can indeed be converted in these extremely energetic particles, requires a better understanding of the resulting state. Such a study would also have to include other species of elementary particles and take into account the induced matter and radiation energy components (see Sec.\ \ref{sec:Balance}) when imposing energy conservation. Superficially, the percentage of space converted into particles at this time should match the percentage of matter observed in the current universe, which is about 5\%. This percentage is much lower than the geometric fractions given above, so even the description of this very early universe can be subjected to constraints based on current observations.

Since the vacuum space is completely homogeneous, this creation of particle-anti-particle pairs will happen everywhere at the same time. Hence, this represents an instantaneous phase transition of the universe from a classical to a quantum state. Because of the limited number of pairs that can be produced, their kinetic energy (or rather their momentum) is enormous at this stage. The bulk of particles is produced later in ordinary collision processes, dispersing the enormous kinetic energy in tandem with the continued expansion of the universe. However, since these particles originate from a very small initial sample, they are strongly correlated and entangled, which may well have an impact on the character of the current distribution of matter and radiation.

At some point the kinetic energy has dissipated to such an extent that the production of particles peters out and annihilation processes become more prominent. The number of particles will then diminish, eventually resulting in a net surplus of quarks and leptons, rather than anti-quarks and anti-leptons. The current formulation may also shed light on the question as to whether this asymmetry is due to a fundamental asymmetry in the interactions, or is of statistical origin.

In the previous discussion we have indicated that it takes a finite time for a particle to be created. We now want to show how this creation time is determined. We consider the case of light quarks, as this case has been previously analyzed in \cite{QuarkDressing} and \cite{QuarkSpatial}. The condition that the spherical vacuum universe stops its expansion at a particular time $t=t_{c}$ when it coincides with the Planck-sized QFT system, yields the following constraint:
\begin{equation}
\label{eq:Creation_time_equation}
 \epsilon\frac{t_s^3}{t^3_{c}}V-\beta_q E= m_q\approx 0;V=\frac{4\pi}{3}r_0^3;Er_0=x_0.
\end{equation}
Here $m_q$ is the resulting positive quark mass and $E$ sets the internal energy scale, which can be shown to be $\textrm{O}(G^{-1/2})$. The radius of the particle $r_0=x_0/E$ is given in terms of the dimensionless number $x_0$. This radius is fixed in the QFT calculation and equals $2.0429\cdots$ for light quarks. The parameter $\beta_q$ expresses the internal potential energy in units of $E$ and is a function of the $\textrm{SU}_{\textrm{n}}$ structure constants of the gauge interactions. The usual coupling constants $\alpha^{(n)}$, which drive the scattering processes in QFT, can in first instance be ignored in the construction of these soliton-like solutions.

Since the particle mass $m_q$ is negligible in comparison to the two $\textrm{O}(G^{-1/2})$ contributions, Eq.\ (\ref{eq:Creation_time_equation}) can be solved by setting $m_q=0$. One then finds:
\begin{equation}
\label{eq:Creation_time}
 t_{c}= \textrm{O}( G^{1/6}/\epsilon^{1/6})=\textrm{O}\left({10^{-23}\, \textrm{s}}\right).
 \end{equation}
Hence, it takes a finite time $t_{c}$ for a real (as opposed to a virtual) elementary particle to be created. This is also the case for the first particles created after the big bang. So the classical epoch lasts for a finite time until the first particles can be created. Different elementary particles will have different creation times $t_{c}$. So, a more complete description will involve different creation times and different particle species which compete for the same space.

The previous discussion has shown the usefulness of the BS-QFT theory of elementary particles in cosmology. In order to provide further support for this methodology, and thus for its use in cosmology, we mention some encouraging results of the first mass calculations in this theory. Unlike typical bound-state calculations, where the mass is an eigenvalue of the Hamiltonian, the mass $m_q$ in Eq.\ (\ref{eq:Creation_time_equation}) is determined by imposing an additional constraint. We assume that the creation time $t_{c}$, which is characteristic for the particle considered, also determines its mass and note that - according to the uncertainty principle - the smallest mass that can be created in the period $t_{c}$ is given by:
\begin{equation}
\label{eq:mass}
 m_q=\frac{\hbar}{2}\frac{1}{t_{c}}.
 \end{equation}
Inserting this in Eq.\ (\ref{eq:Creation_time_equation}) turns it into a self-consistent equation, with a solution in the range:
\begin{equation}
\label{eq:mass_size}
 m_q =\textrm{O}( \epsilon^{1/6}/G^{1/6})\approx 40 \,\textrm{MeV}.
 \end{equation}
This mass range lies right in the particle physics mass domain, despite being derived from the gravitational constant $G$ and the cosmological constant $\Lambda$ (or $\epsilon$). So, this result provides an important bridge between particle physics, general relativity and cosmology, as will be further investigated in Sec.\ \ref{sec:Big_Numbers}.

The first concrete application of this BS-QFT theory used only $\textrm{SU}_{\textrm{3}}$-interactions and produced a ground state mass of about $3.2\, \textrm{MeV}$  \cite{QuarkDressing}. This value (and a slightly larger value found in a more recent study \cite{QuarkSpatial}) lies right in the empirical range for light quarks $3.8 \pm 0.8\,\textrm{MeV}$, known from lattice gauge calculations \cite{Lattice Quark masses}. This is a surprisingly good result, emerging as it does from the cancellation between two numbers which are 22 orders of magnitude in excess of the mass itself. This mechanism also puts into question the need for the - currently popular - naturalness hypothesis.

So far, the masses of the more massive quarks come out too low, as they still lie in the MeV energy range. But this may be due to the fact that the Higgs field, which is known to play a big role for heavier quarks, has not yet been included. There are preliminary indications that the Higgs field does not couple to the ground-state solution, which might explain why the average light quark mass is better predicted than the heavier ones. Since the current cosmological theory plays an important role in this BS-QFT theory (and vice versa), the successful prediction of the average light quark mass can also be seen as support for the current cosmological theory.

\section{Inclusion of matter and radiation}
\label{sec:Matter_Rad}
We now consider the consequences of the presence of matter and radiation in the universe. We assume that vacuum (dark) energy keeps dominating the energy of the universe in this more mature phase of its development, aided by the present consensus that dark energy dominates the energy content of the universe \cite{Kowalski}. This means that the additional components can be handled perturbatively, suggesting the following expansions:
\begin{eqnarray}
 \label{eq:h_pertubation}
 \begin{aligned}
g_{\mu\nu}=&g(t)\left[\eta_{\mu\nu}+h_{\mu\nu}(x)\right]
 \\
g^{\mu\nu}=&g^{-1}(t)\left[\eta^{\mu\nu}+h^{\mu\nu}(x)\right],
 \end{aligned}
\end{eqnarray}
where $h_{\mu\nu}$ represents the first order correction due to the presence of matter or radiation. Apart from the conformal factor in front, this expansion corresponds to the usual weak gravity approximation (see \cite{MTW}, p. 435). We also follow \cite{MTW} in adopting the notation $h^{\mu\nu}$, although $h_{\mu\nu}$ is not a covariant tensor.
We now divide the energy-momentum tensor in a vacuum and a quantum part, the latter consisting of a matter and radiation part:
\begin{equation}
\label{eq:Tuv}
T_{\mu\nu}=T_{\mu\nu}^{(vac)}+T_{\mu\nu}^{(qu)}=-\epsilon g_{\mu\nu}+T_{\mu\nu}^{(m)}+T_{\mu\nu}^{(rad)}.
\end{equation}
We can use the completeness condition $g_{\mu\alpha} g^{\alpha\nu}=\delta_{\mu}^{\;\nu}$ to derive a first-order expression for
$h^{\mu\nu}$ in terms of $h_{\mu\nu}$:
\begin{equation}
\label{eq:h_inverse}
h^{\mu\nu}=-\eta^{\mu\mu}h_{\mu\nu}\eta^{\nu\nu}-\eta^{\mu\mu}h_{\mu\alpha}h^{\alpha\nu}\approx -\eta^{\mu\mu}h_{\mu\nu}\eta^{\nu\nu},
\end{equation}
where in the last step we ignored the term quadratic in $h_{\mu\nu}$. This expression is identical to the one found in the weak-gravity case, as the conformal factors $g$ and $g^{-1}$ cancel. 	

Note that the different treatment of classical vacuum energy (non-perturbatively) and the matter and radiation energy (perturbatively) does not just reflect the presumed dominance of vacuum energy, but also reflects the difference between classical (invertible) c-number quantities like $g(t)$; and (non-invertible) quantum q-number fields like $h_{\mu\nu}$. This distinction is blurred in cosmology because we usually deal with c-number-like expectation values of quantum operators. However, this distinction can be relevant in the early universe, when quantum physics and general relativity overlap.

Now we write down the Einstein equation for $h_{\mu\nu}$, cancelling out the vacuum terms arising from the vacuum energy $-\epsilon g_{\mu\nu}$. We obtain the equation:
\begin{eqnarray}
 \label{eq:h_equation}
 \begin{aligned}
&\frac{2}{t}\left(\partial_{\mu}h_{\nu 0}+\partial_{\nu}h_{\mu 0}-\partial_0 h_{\mu\nu}\right)-\frac{4}{t}\eta_{\mu\nu}\partial^{\alpha}(h_{0 \alpha}-\frac{1}{2}\eta_{0 \alpha}h)
\\
&-\frac{6}{t^2}h_{00}\,\eta_{\mu\nu}+\partial_{\mu}\partial^{\lambda}h_{\nu\lambda}
+\partial_{\nu}\partial^{\lambda}h_{\mu\lambda}-\partial_{\nu}\partial_{\mu}h+
\\
&+\eta_{\mu\nu}\left(\Box h-\partial^\alpha\partial^\beta h_{\alpha\beta}\right)-\Box h_{\mu\nu}=16\pi G\,T_{\mu\nu}^{(qu)},
 \end{aligned}
\end{eqnarray}
where $h$ is defined by:
\begin{equation}
\label{eq:h_definition}
h=\eta^{\alpha\beta}h_{\alpha\beta},
\end{equation}
and $\Box = \eta^{\alpha\beta}\partial_{\alpha}\partial_{\beta}$. Dropping the vacuum terms (terms with an explicit time dependence) yields the usual perturbative – weak gravity – equations (see for example: \cite{MTW}, Ch.\ 18; \cite{Weinberg}, Eq.\ (10.1.4) and \cite{Carroll}). This ability to transform smoothly to these well-studied equations can be seen as another advantage of the conformal metric.

By introducing the bar-notation (see \cite{MTW}, Eq.\ (18.6)):
\begin{equation}
\label{eq:bar_notation}
\bar{h}_{\mu\nu}=h_{\mu\nu}-\frac{1}{2} \eta_{\mu\nu}h;\bar{T}_{\mu\nu}^{(qu)}=T_{\mu\nu}^{(qu)} -\frac{1}{2}g_{\mu\nu}T^{(qu)},
\end{equation} 						
this equation can be simplified slightly:
\begin{eqnarray}
 \label{eq:h_equation_reduced}
 \begin{aligned}
&\frac{2}{t}\left(\partial_{\mu}h_{\nu 0} +\partial_{\nu}h_{\mu 0}-\partial_0 h_{\mu\nu}
+\eta_{\mu\nu}\partial^{\alpha}\bar{h}_{0 \alpha}\right)+\frac{6}{t^2}\eta_{\mu\nu} h_{00}
\\
&+ \partial_{\mu}\partial^{\lambda}\bar{h}_{\nu\lambda}
+\partial_{\nu}\partial^{\lambda}\bar{h}_{\mu\lambda}
-\Box h_{\mu\nu}=16\pi G\,\bar{T}_{\mu\nu}^{(qu)},
 \end{aligned}
\end{eqnarray}
This agrees with Eq.~(10.1.4) in \cite{Weinberg} if the vacuum terms are omitted. In \cite{MTW}, Eq.~(18.7) this equation is expressed in
terms of $\Box \bar{h}_{\mu\nu}$ and $T_{\mu\nu}^{(qu)}$.

The first step towards solving these equations is to determine the constraints on the energy-momentum tensor $T_{\mu\nu}^{(qu)}$ set by the vanishing of the covariant divergence:
\begin{equation}
\label{eq:Divergence_condition}
\nabla_{\mu} T^{\mu(qu)}_{\:\nu} =\partial_{\mu} T^{\mu(qu)}_{\:\nu} -\Gamma^{\lambda}_{\mu\nu}T^{\mu(qu)}_{\:\lambda}+\Gamma^{\mu}_{\mu\lambda}T^{\lambda(qu)}_{\:\nu}=0,
\end{equation}
where we have dropped the vacuum contribution, as $\nabla_{\mu}T^{\mu (vac)}_{\nu}$ is automatically zero. In our perturbative treatment the connection coefficients are determined by the vacuum metric, so that the equation has the same form as in the vacuum case:
\begin{equation}
\label{eq:Divergence_vacuum}
\partial_{\mu}T^{\mu(qu)}_{\nu}=\frac{3}{t}T^{0(qu)}_{\nu}-\frac{1}{t}\left(\delta_{\nu 0}T^{(qu)}+\eta_{\nu\nu}T^{0(qu)}_{\nu}\right).
\end{equation}

In the matter case the energy-momentum tensor only has a 00 - component, so that:
\begin{eqnarray}
\label{eq:Divergence_matter}
 \begin{aligned}
\partial_0 &T^{0(m)}_0=\frac{3}{t}T^{0(m)}_0-\frac{1}{t}\left(T^{(m)} -T^{0(m)}_0\right)=
\\
& =\frac{3}{t}T^{0(m)}_0 \rightarrow T^{0(m)}_0\propto t^3.
 \end{aligned}
\end{eqnarray}
We now model the matter energy tensor subject to this condition, while taking into account the essentially point-like nature of matter in the vacuum dominated universe:
\begin{equation}
\label{eq:T_matter}
  T^{0(m)}_0(x)=-\rho^{(m)}(x)=-\sum\limits_i\frac{M_i}{g(t)^{3/2}}\delta^{(3)}(\vec{x}-\vec{x_i}).
  \end{equation}
In general relativity the $\delta$-function must be accompanied by the inverse Jacobian, which in 3 dimensions reads $g(t)^{-3/2}$ (see \cite{Weinberg} for the corresponding 4-dimensional case: Eq.\ (5.2.13), p.\ 125). This leads directly to the desired $t^3$ behavior. The reason for using the explicit representation in terms of masses is that in an expanding universe they have a clearer physical meaning than the density. Note that $M_i$ may include the kinetic energy and nuclear binding, as was stated in \cite{MTW}, p.\ 140.

This improved approach to the treatment of matter in cosmology is not available in the usual FRW framework, where only time dependence can be considered in accordance with the cosmological principle and the ideal fluid assumption. However, matter in the universe is not distributed homogeneously; and by accounting for the spatial dependence in Eq.\ (\ref{eq:T_matter}), we can also include the important effect of spatial derivatives in the Einstein equations. We can always return to the usual spatially independent densities by carrying out spatial averages.

The matter energy in the volume $V(t)$ then equals:
\begin{equation}
\label{eq:Etot_matter}
  E^{(m)}_{V(t)}=-\int_{V(t)}d^3x g(t)^{3/2}T^{0(m)}_0(x)=\sum\limits_{i\in {V(t)}} M_i.
\end{equation}
This energy is conserved if matter co-moves with the expansion of space, which is the case unless matter is bound by quantum forces, such as in molecules, atoms and nuclei. In contrast to the vacuum case the volume is no longer arbitrary and must be large enough for the cosmological principle to be applicable, so that we can subsequently average over space to define the spatially independent quantities needed in cosmology.

The barred source term which was defined in Eq.\ (\ref{eq:bar_notation}) can now be written to lowest order as:
\begin{equation}
\label{eq:T_source}
  \bar{T}_{\mu\nu}^{(m)}=\frac{1}{2}\delta_{\mu\nu}T_{00}^{(m)}.
\end{equation}
Eq.\ (\ref{eq:h_equation_reduced}) then suggests that the matter solution $h_{\mu\nu}^{(m)}$ may well have a diagonal form, as well. So we try:
\begin{equation}
\label{eq:hm_model}
  h_{\mu\nu}^{(m)}(x)=\delta_{\mu\nu}\;h^{(m)}(x),
\end{equation}
and find that Eq.\ (\ref{eq:h_equation_reduced}) does indeed yield a solution with $h^{(m)}(x)$ given by:
\begin{eqnarray}
\label{eq:hm_solution}
 \begin{aligned}
 h^{(m)}(x)& =2G\frac{t}{t_s}\sum\limits_i \frac{M_i}{|\vec{x}-\vec{x}_i|}\theta(ct-|\vec{x}-\vec{x}_i|)
\\
 \equiv 2G\frac{t}{t_s}& \int d^3x' \frac{\rho^{(m)}(\vec{x}')}{|\vec{x}-\vec{x}'|}\theta(ct-|\vec{x}-\vec{x}'|).
 \end{aligned}
 \end{eqnarray}
Note that the spatial boundaries of the integral are not defined by $V(t)$, but by the $\theta$-function which ensures that only matter that is in causal contact with $(\vec{x},t)$ contributes, thereby ensuring that the sum or integral remains finite. Since the matter at large distance from $\vec{x}$ refers to a much earlier time than $t$, the associated density may be different from the one at time $t$. For now we ignore this complication, but in our discussion of induced (dark) matter terms, we will come back to it.

Both $T_{00}^{(m)}=g_{00}T^{0(m)}_0$ and $h^{(m)}$ are now linear in $t$. This is the explicit (dynamic) time dependence which is relevant for the solution of the Einstein equations. The expansion of the universe counters this behavior, so that in the co-moving representation - which is relevant for the calculation of the energy integrals - there is no residual time dependence and the expressions turn into the usual spatial Newtonian expressions:
\begin{equation}
\label{eq:hm_wiggle}
  \tilde{h}^{(m)}(\tilde{x})=2G\int d^3\tilde{x}'
  \frac{\tilde{\rho}^{(m)|}(\vec{\tilde{x}}')}{|\vec{\tilde{x}}-\vec{\tilde{x}}'|}\theta(ct_s-|\vec{\tilde{x}}-\vec{\tilde{x}}'|),
  \end{equation}
where $\vec{\tilde{x}}$ was defined in Eq.\ (\ref{eq:co_moving}). The density can similarly be expressed in terms of co-moving variables:
\begin{equation}
\label{eq:rho_matter_avrg}
  \tilde{\rho}^{(m)}(\vec{\tilde{x}})=\sum\limits_i M_i \delta^{(3)}(\vec{\tilde{x}}-\vec{\tilde{x}}_i);
  \langle \tilde{\rho}^{(m)}\rangle = \frac{1}{\tilde{V}}\sum\limits_{i\in \tilde{V}} M_i.
\end{equation}

These formulae also hold for local weak gravity systems, such as planetary systems. Hence, in terms of co-moving variables such systems can be described using the usual Newtonian expressions as if there is no expansion. This may explain why planetary systems remain stable despite the expansion of the universe. Of course, higher-order local corrections are not included here, as we only expand up to first order terms. 


One can also calculate the spatial average of $\tilde{h}^{(m)}(x)$ and finds:
 \begin{equation}
\label{eq:h_matter_avrg}
  \langle \tilde{h}^{(m)}\rangle =\frac{3}{2} \frac{\langle \tilde{\rho}^{(m)}\rangle}{\epsilon}.
\end{equation}
This is a simple - intuitively appealing - result for the average effect of the matter density on the metric. In the next section we will see that this leads to an induced energy contribution, whose magnitude is $9/4$ times the normal matter contribution.


Now we consider the radiation case. In cosmology the energy-momentum tensor for radiation is modelled as a diagonal tensor with zero trace:
\begin{equation}
\label{eq:T_rad}
  T_{\mu}^{\nu(rad)}(x)=\rho^{(rad)}(t)\left(-\frac{4}{3}\delta_{\mu 0}\delta_{\nu 0}+\frac{1}{3}\delta_{\mu}^{\nu}\right).
\end{equation}
The divergence condition Eq.\ (\ref{eq:Divergence_vacuum}) implies that $T_0^{0(rad)}$ behaves like $t^4$, which suggests the following parametrization for the radiation density in the vacuum background metric:
\begin{equation}
\label{eq:rho_rad}
  \rho^{(rad)}(t)=\frac{1}{g^2(t)}\frac{1}{V}\sum\limits_{j\in V} p_0^{(j)}.
\end{equation}
Here $p_0^{(j)}$ is the energy of the $j^\textrm{{th}}$ photon inside the volume $V$ at time $t$. Since we use the spatially averaged expression, $V$ must be big enough for the cosmological principle to apply.

These expressions are consistent with conservation of energy. To prove this one replaces
the photon four-momentum $p_{\mu}^{(j)}$ by its co-moving partner $\tilde{p}_{\mu}^{(j)}$:
\begin{equation}
\label{eq:co_pmu}
 \tilde{p}_{\mu}^{(j)}=\frac{t}{t_s}p_{\mu}^{(j)}.
\end{equation}
which is constant under the dominant linear expansion of the universe. The radiation energy can now be written as follows:
\begin{eqnarray}
\label{eq:E_rad}
\begin{aligned}
&E^{(rad)}_{V} = -\int\limits_{V(t)}d^3x \, g^{3/2}(t)T_0^{0,(rad)}(x)=
\\
&=\frac{t}{t_s}\sum\limits_{j \in {V(t)}} p_0^{(j)}=\sum\limits_{j\in {V(t)}} \tilde{p}_0^{(j)}\equiv\sum\limits_{j\in \tilde{V}} \tilde{p}_0^{(j)}.
\end{aligned}
\end{eqnarray}
This energy is conserved as the $\tilde{p}_{\mu}^{(j)}$ are constant. We can also relate these momenta to their known spectroscopic value at creation: $\hat{p}_0^{(j)}$. If the photon was created at $t=t^{(j)}$ then:
\begin{equation}
\label{eq:p_creation}
p_{\mu}^{(j)}=\frac{t^{(j)}}{t}\hat{p_{\mu}}^{(j)}.
 \end{equation}
Both $\tilde{p}_{\mu}^{(j)}$ and $\hat{p}_{\mu}^{(j)}$ are constant and related by the factor $t^{(j)}/t_s$.

Note that the product of momentum and space-time variables is invariant under the transition to the co-moving representation:
\begin{equation}
\label{eq:Scaling_relation}
  x\cdot p=\tilde{x}\cdot \tilde{p}\; .
\end{equation}
This complementarity facilitates the use of the co-moving representation in an expanding universe and in quantum physics, as plane waves - which play an important role in the expansions of QFT amplitudes - have the same form in both representations:
$\exp(\textrm{i}\, x\cdot p)=\exp(\textrm{i}\, \tilde{x}\cdot \tilde{p})$.

We can solve Eq.\ (\ref{eq:h_equation}) by setting:
 \begin{equation}
\label{eq:hij}
  h_{\mu\nu} \rightarrow h^{(rad)}_{\mu\nu}=\delta_{\mu\nu}\left[\delta_{\mu 0}h^{(rad)}_{00}+\bar{\delta}_{\mu 0}h^{(rad)}\right],
\end{equation}
and find the solution:
\begin{eqnarray}
\label{eq:Rad_solution}
\begin{aligned}
h^{(rad)}_{00}  -4h^{(rad)} =\frac{8\pi G}{3}t^2T^{(rad)}_{00}=\frac{\rho^{(rad)}(t)}{\epsilon}.
\end{aligned}
\end{eqnarray}
The reason that we did not find separate solutions for $h^{(rad)}_{00}$ and $h^{(rad)}$ is that after inserting Eq.\ (\ref{eq:hij}) into Eq.\ (\ref{eq:h_equation}), the two resulting equations are dependent.

In the co-moving representation we have to take into account the time dependence of $V$ and $p_0^{(j)}$, so that the radiation density becomes constant:
\begin{equation}
\label{eq:rho_rad_comoving}
\rho^{(rad)}(t) \rightarrow \tilde{\rho}^{(rad)}=\frac{1}{\tilde{V}} \sum\limits_{j\in \tilde{V}} \tilde{p}_0^{(j)}.
\end{equation}

The non-uniqueness of the solutions in Eq.\,(\ref{eq:Rad_solution}) also occurs in the weak-gravity equations. MTW \cite{MTW}, Eq.\,(18.8a) and Weinberg \cite{Weinberg}, Eq.\,(10.1.9) resolve this ambiguity, by imposing the condition:
\begin{equation}
\label{eq:gauge_condition}
\partial^{\nu}\bar{h}_{\mu\nu}=0.
\end{equation}
In weak gravity this condition reduces the differential equations to Klein-Gordon form, suggesting a correspondence to the Maxwell equation and the gauge freedom therein. In our formulation Eq.\,(\ref{eq:gauge_condition}) implies:
\begin{equation}
\label{eq:h00_hrad}
  h_{00}^{(rad)}=-3h^{rad},
\end{equation}
so that the co-moving solutions become:
\begin{equation}
\label{eq:explicit_solution_rad}
\tilde{h}^{(rad)}=-\frac{1}{7}\frac{\tilde{\rho}^{(rad)}}{\epsilon};
\tilde{h}_{00}^{(rad)}=\frac{3}{7}\frac{\tilde{\rho}^{(rad)}}{\epsilon}.
\end{equation}
In \cite{FOS_Greben} we applied another constraint, by assuming that $h_{\mu\nu}^{(rad)}$ is proportional to $T_{\mu\nu}^{(rad)}$, so that $h_{00}^{(rad)}=3h^{(rad)}$. The co-moving solution then reads:
\begin{equation}
\label{eq:explicit_solution_rad_JMG}
\tilde{h}^{(rad)}=-\frac{\tilde{\rho}^{(rad)}}{\epsilon};\;
\tilde{h}_{00}^{(rad)}=-3\, \frac{\tilde{\rho}^{(rad)}}{\epsilon},
\end{equation}
which is a factor 7 larger in magnitude than the first one, with $\tilde{h}_{00}^{(rad)}$ even changing sign. Since the different solutions lead to different contributions to the total energy, it is important that this ambiguity is resolved in the future. One possibility is that the ambiguity is only present in the perturbative linear equations and will go away as soon as higher-order terms are included.


Note that both in the matter and the radiation case, $\tilde{h}$ is proportional to $\tilde{\rho}/\epsilon$. This result is intuitively appealing and also plays an important role in the upcoming discussion of the composition of the total energy. These ratios resemble those in the standard $\Lambda$CDM model, with the vacuum energy density replaced by the so-called critical density (see discussion at the end of Sec.\ \ref{sec:Balance}). This similarity gives us confidence that in both theories the same parameters must be fit to explain the cosmological observations, so that their predictions may be much closer than their different bases would have us to expect.


\section{Energy Balance and the Expansion of the Universe}
\label{sec:Balance}
In Sec.\ \ref{sec:Analysis} we discussed how the energy in the universe could have been produced in the big bang without violating energy conservation through the action of a negative pressure term in the vacuum energy tensor. We then postulated that after the big bang this energy is conserved, initially in the form of vacuum energy, and later as a mixture of different forms of energy.

Our description of the creation epoch in Sec.\ \ref{sec:Creation} shows that the initial classical de Sitter space will persist for a finite time. During this classical epoch, the energy in a co-moving volume $\tilde{V}$ is given by:
 \begin{equation}
\label{eq:classical}
 E^{(tot)}_{\tilde{V}}=\epsilon \frac{t_s^3}{t^3}V(t)=\epsilon \tilde{V}\equiv \epsilon V_s;\, t<t_{cl},
\end{equation}
where $t_{cl}$ marks the end of the classical epoch and the moment when the first real (as opposed to virtual) elementary particles can be created.
In this phase the size of $\tilde{V}$ plays no role, as space is completely homogeneous.

Once this classical period has come to an end, it is important to consider a big enough volume $\tilde{V}$, so that the cosmological principle and the associated statistical arguments can be applied. Considerable changes in the composition of the universe will now take place, however, the total energy in a (sufficiently large) co-moving volume will remain the same, as the difference between the energy moving in and out of the volume through radiation or collisions is expected to be negligible. During these changes we expect vacuum energy to remain the dominant energy component, with its current percentage being estimated at 71.3\% \cite{Bennett}.

Hence, in the analysis of the total energy we can continue with our perturbative approach and treat terms like $T^{(qu)}$ or $h$ only to first order. Terms linear in $T^{(qu)}$ lead to the  conserved matter and radiation contributions which were discussed in the previous section. The terms proportional to $h$ modify the original vacuum energy integral and can be shown to be proportional to the direct matter and radiation terms. These novel contributions may yield a possible explanation for dark matter.

The total energy in the presence of matter and radiation can now be written as follows:
 \begin{eqnarray}
 \label{eq:Energy_1}
 \begin{aligned}
 E^{(tot)}_{\tilde{V}}=\epsilon \tilde{V}=-\int\limits_{V(t)}d^3x|g_{ij}|^{1/2}[T^{0,(vac)}_{\ 0}+T^{0,(qu)}_{\ 0}]=
 \\
 =\int\limits_{V(t)}d^3x\frac{t^3_s}{t^3}[1+\tilde{h}^{(m)}+\tilde{h}^{(rad)}]^{3/2}[\epsilon+\tilde{\rho}^{(m)}+\tilde{\rho}^{(rad)}],
 \end{aligned}
\end{eqnarray}
where we have used the fact that the spatial components of the diagonal matrix $h_{ij}$ are identical.
Keeping only terms up to first order in $\tilde{h}$ and $\tilde{\rho}$, we get:
\begin{eqnarray}
\label{eq:Energy_2}
\begin{aligned}
&E^{(tot)}_{\tilde{V}}= \epsilon \tilde{V} \approx \epsilon\int\limits_{V(t)}d^3x\frac{t^3_s}{t^3}+
\\
+&\int\limits_{V(t)}d^3x\frac{t^3_s}{t^3}[\tilde{\rho}^{(m)}+\tilde{\rho}^{(rad)}+\frac{3}{2}\epsilon(\tilde{h}^{(m)}+h^{(rad)})].
\end{aligned}
\end{eqnarray}
Since the co-moving matter and radiation energy densities are constant, and $\tilde{h}^{(m)}$ and $h^{(rad)}$ are proportional to these densities, all terms in Eq.\ (\ref{eq:Energy_2}) remain constant if $V(t)\sim t^3$. However, the reference volume $\tilde{V}=V_s$ must now change (be reduced) to accommodate the new terms, so we write:
\begin{equation}
\label{eq:Vt}
V(t)=\frac{t^3}{t^3_s}\tilde{V}_t;\, \tilde{V}_t=\tilde{V}\; \textrm{for}\;  t<t_{cl}.
\end{equation}
We then get:
\begin{eqnarray}
 \label{eq:Energy_4}
 \begin{aligned}
 \epsilon \tilde{V} =
 \tilde{V}_t&\left[\epsilon +\tilde{\rho}^{(m)}\left(1+\frac{9}{4}\right)+\tilde{\rho}^{(rad)}\left(1-\frac{3}{14}\right)\right],
 \end{aligned}
\end{eqnarray}
where we have inserted the average value of $\tilde{h}^{(m)}$ from Eq.\ (\ref{eq:h_matter_avrg}) and the solution in Eq.\ (\ref{eq:explicit_solution_rad}) for $\tilde{h}^{(rad)}$.

The time dependence of the co-moving volume $\tilde{V}$ is due to the fact that the relative contributions of the vacuum, matter and radiation energy change over time because of quantum transitions. To model this implicit time dependence of $\tilde{V}$ and the densities requires a physical analysis of the prevailing processes in subsequent epochs. As noted before, theses implied time dependencies do not enter explicitly in the Einstein equations; so they play no dynamical role.

In the $\Lambda$CDM model general relativity is attributed magical powers, as the FRW solutions for the scale factor determine which densities prevail at certain times. However, to conclude that radiation dominates in the early universe on the basis of the idealized FRW solutions for radiation ($\sim a(\tau)^{-4}$) and matter ($\sim a(\tau)^{-3}$) densities, makes no physical sense, as such dominance should be determined by quantum theory and the known gauge forces, with general relativity only acting as a constraint. Radiation is caused by the decay of particles or their scattering and annihilation, so how can radiation dominate before matter is present?



Let us now discuss the extra components in Eq.\ (\ref{eq:Energy_4}). Could the extra matter term with coefficient 9/4 be an explanation for the phenomenon of dark matter? The fact that it depends on the presence of matter is suggestive thereof. Analyzing the original $h^{(m)}(\vec{x},t)$ in Eq.\ (\ref{eq:hm_solution}) shows that the matter that contributes to the integral is centered around $\vec{x}$, further reinforcing the dark matter interpretation. However, the large magnitude of this integral being proportional to $1/\epsilon$, is due to its slow convergence and the importance of the contribution of distant matter, which weakens the dark matter hypothesis. That the factor 9/4 is considerably smaller than the empirical ratio 5.2 \cite{Bennett} of dark over normal matter, should be less of a concern, as there are various refinements which could lead to a larger number. For example, the assumption that the matter density was larger in the past, would immediately enhance this factor. Clearly, further analysis - also involving tests against the galaxy rotation curves \cite{Gaugh}, \cite{Paolo} - is needed to come to a firm conclusion about this dark matter hypothesis.

The induced radiation term has a coefficient -3/14 if the solution in Eq.\ (\ref{eq:explicit_solution_rad}) is used. This extra term reduces, rather than enhances, the contribution of radiation to the total energy. Its magnitude depends on the way the ambiguity in Eq.\ (\ref{eq:Rad_solution}) is handled. The other solution in Eq.\ (\ref{eq:explicit_solution_rad_JMG}) leads to a coefficient -3/2 and even inverts the overall sign. It is not clear how this term would reveal itself in direct observations. However, it plays an important role in explaining the recent acceleration of the universe, as we show next. So this may be a way to confirm its presence and determine its magnitude.

If matter is converted into radiation, then the induced matter term decreases in size, while the extra radiation term increases in size, but with opposite sign. Hence both cross terms lead to a reduction of the total energy when matter turns into radiation:
\begin{equation}
\label{eq:Energy_reduction}
\triangle E^{(tot)}_{\tilde{V}_t}=\tilde{V}_t\left(\frac{9}{4} \triangle\tilde{\rho}^{(m)}-\frac{3}{14} \triangle \tilde{\rho}^{(rad)}\right),
\end{equation}
where $ \triangle \tilde{\rho}^{(rad)}=-\triangle \tilde{\rho}^{(m)}$, as energy is conserved in a quantum transition. This loss in energy would be even bigger if the factor $9/4$ is replaced by the empirical dark matter value of 5 and if instead of 3/14 we use the value 3/2 in the radiation case. The loss in energy must now be compensated by a corresponding gain in vacuum energy through an acceleration of the linear expansion, i.e.\ an increase in $\tilde{V}_t$:
\begin{equation}
\label{eq:Energy_volume}
\frac{\triangle \tilde{V}_t}{\tilde{V}_t}=-\frac{\triangle E^{(tot)}_{\tilde{V}_t}}{E^{(tot)}_{\tilde{V}_t}}.
\end{equation}
By developing physical models for the time dependence of the matter and radiation densities one can attempt to quantify these estimates and test their validity against the observed acceleration. This may also help to resolve the ambiguity in the choice of the radiation solution.

The densities in Eq.\ (\ref{eq:Energy_4}) can also be written as ratios over the dominant vacuum energy density $\epsilon$. They then resemble the ratios occurring in the standard $\Lambda$CDM model, except that the latter are defined with respect to the critical density $\rho^{(0)}_c$. This critical density, which is supposed to mark the boundary between an open and closed universe, is defined as follows (see \cite{Weinberg}, p.\ 475):
 \begin{equation}
\label{eq:Critical_density}
\rho^{(0)}_c=\frac{3H_0^2}{8\pi G},
\end{equation}
where $H_0$ is the current Hubble constant. In our conformal theory we have:
\begin{equation}
\label{eq:Hubble_t}
H(t)=\frac{\dot{a}_{conf}(t)}{a_{conf}(t)}=\frac{1}{t}.
\end{equation}
For a proper (co-moving) observer we thus have:
\begin{equation}
\label{eq:t_s}
H_0=\frac{t}{t_s}\frac{1}{t}=\frac{1}{t_s}.
\end{equation}
This expression actually agrees with the FRW result if the standard (increasing) exponential FRW scale factor is used in Eq.\ (\ref{eq:scale_factor}). It shows how totally different theories can still yield the same predictions, complicating their independent validation.




Inserting this result in the critical density we find that it equals the vacuum energy density $\epsilon$:
\begin{equation}
\label{eq:Crit_density_vacuum_density}
 \rho^{(0)}_c=\frac{3}{8\pi G}\frac{1}{t_s^2}=\epsilon.
\end{equation}
Hence, in our formulation there is no critical density and the fact that the current density of the universe is so close to the so-called critical density (known as the coincidence problem) is simply a consequence of the dominance of vacuum energy in the universe. There is no theoretical constraint on the value of the vacuum energy density $\epsilon$, and it must simply be determined through observation, as it (or rather the cosmological constant) is one of the basic (dimensionful) constants of nature.

As stated above, another consequence of the similarity between the ratios $\tilde{\rho}/\epsilon$ and $\rho/\rho_c$ could be that the cosmological predictions of the $\Lambda$CDM model and the current theory are much more similar than the differences in the underlying frameworks would suggest; after all in both cases the parameters are designed to reproduce the same cosmological observations.

\section{Big Numbers}
\label{sec:Big_Numbers}

In further support of the conformal cosmological theory we conclude with the discussion of a puzzle which has occupied the minds of great physicists like Eddington \cite{Eddington}, Dirac \cite{DiracLarge} and Weinberg \cite{Weinberg}. This is the question of the so-called \emph{big numbers}, the fact that many ratios between cosmological parameters and particle physics parameters are multiples of $10^{20}$ or $10^{40}$ and/or their inverses (see \cite{MTW}, p. 1216). Dirac \cite{DiracLarge} and others suspected that these outcomes were not mere coincidences and required a physical explanation.

We follow Weinberg (\cite{Weinberg}, p.\ 619-620) in discussing some of these coincidences. The first one is that the ratio of the Hubble constant and $G$ can be related to the typical mass of an elementary particle:
\begin{equation}
\label{eq:Hubble_G}
 (H_0/G)^{1/3}=\mathcal{O}(60\, \textrm{MeV}),
 \end{equation}
where Weinberg used a Hubble time of $H_0^{-1}=10^{10}\,\textrm{years}$. Since $H_0^{-1}=t_s$ in the current theory
(see Eq.\  (\ref{eq:t_s})), and $(t_sG)^{-1/3}=(4\pi \epsilon/3 G)^{1/6}$, Eq.\ (\ref{eq:Hubble_G}) can directly be related to Eq.\ (\ref{eq:mass_size}).

The other accidental relationship discussed by Weinberg is (see \cite{Weinberg}, Eq.\ (16.4.3)):
\begin{equation}
\label{eq:Baryon}
 G\,n_0\,m_p\approx H_0^2,
 \end{equation}
where $n_0$ is the baryon number density. Since $n_0\, m_p\equiv \langle \tilde{\rho}^{m}\rangle$, and $\langle\tilde{\rho}^{m}\rangle$ is comparable to $\epsilon$, we find that Eq.\ (\ref{eq:Baryon}) also follows from our expression for the Hubble constant in Eq.\ (\ref{eq:t_s}).

\section{Summary}
\label{sec:Summary}
In this paper we propose a new cosmological theory in which the big bang and the subsequent expansion of the universe are controlled by energy conservation and the cosmological constant and its associated vacuum energy.

The universe starts as a classical de Sitter space filled with pure vacuum energy. The effective energy density is given by $\epsilon t_s^3/t^3$, where $\epsilon$ is proportional to the cosmological constant and $t$ is the conformal time. To counter the decrease in energy the universe (must) expand like $t^3$. Until the creation of the first elementary particles starts, this classical state persists and maintains its zero entropy and complete homogeneity.

The analysis of this initial classical universe also reveals that the usual interpretation of the FRW scale factor, namely that it describes the expansion of the universe, is incorrect, and that the true cause of this expansion lies in energy conservation. Hence, the evolution of the universe is not controlled by FRW equations and the associated scale factors, but by the imposition of energy conservation in the presence of the dominant vacuum (dark) energy. So, instead of describing the universe as a superposition of idealized (vacuum, matter or radiation) universes, subsequent epochs are now described as a logical continuation of prior ones, taking into account the physical processes that are possible in the particular phases.

The first such transition takes place at the end of the classical epoch. In the absence of pre-existing particles, a unique physical process is required to create the first elementary particles. Once the vacuum energy density has fallen to the Planck level $\sim G^{-2}$, creation becomes possible by the direct conversion of vacuum energy into particle-anti-particle pairs. To describe this process, which happens at specific times for different particle species, use is made of a BS-QFT theory where elementary particles are treated as finite Planck-sized systems. General relativity and vacuum energy also enter in this theory, making it an ideal theoretical laboratory for testing the unification of these three basic disciplines in a Planck environment.

The number of particles that can be created directly from the vacuum is limited by the available space. The enormous kinetic energy of these particles is subsequently distributed through more conventional collision processes by the creation of further generations of particles, all subject to the space that becomes available with the advance of time. This phase comes to an end when the universe has cooled to such an extent that the creation of particles peters out and annihilation processes become prominent. The fact, that the matter and radiation which is present after this period originate from a very small set of initial particles, may well introduce long-range correlations which leave an imprint on the matter and radiation distributions which are currently observed. This is one of the questions which calls for further investigation.





Because vacuum (dark) energy is assumed to continue its dominance in the later universe, matter and radiation can be treated perturbatively. This allows for a refinement of the cosmological treatment, as we can now take into account the essentially pointlike nature of the matter distribution in the universe without abandoning the cosmological principle. It also allows us to integrate the global linear expansion of the universe in the weak gravity description of local gravitational problems.

The perturbative approach also allows us to write the total energy as a sum of vacuum, matter and radiation components plus induced matter and radiation terms, which are a consequence of the change in the vacuum metric. The induced terms, which can possibly be interpreted as dark matter (and radiation), also provide an explanation for the recent acceleration of the universe, now that the common explanation (the cosmological constant) no longer applies. However, the current uncertainty in the size of the induced radiation term must be resolved before firmer predictions can be done.

In summary, the new cosmological theory can explain many properties of the actual universe with only two fundamental constants: the gravitational and cosmological constant. The finding that the density of the universe is so close to the critical density is no longer a mystery, as this density simply equals the vacuum energy density. The theory also provides good quantitative estimates of the Hubble constant and the age of the universe. The model also plays an essential role in the theoretical prediction of the masses of (light) elementary particles, emphasizing the importance of unifying QFT, general relativity and cosmology in a common framework.
\begin{acknowledgments}
The author acknowledges extensive discussions with Prof.\ Piet Mulders (VU Amsterdam and NIKHEF).
He also acknowledges earlier communications with Prof.\ George Ellis (UCT, Cape Town)
\end{acknowledgments}

\end{document}